\documentclass[aps,prd,preprint,superscriptaddress,tightenlines,nofootinbib]{revtex4}

\usepackage{graphicx}% Include figure files
\usepackage{dcolumn}% Align table columns on decimal point
\usepackage{bm}% bold math

\newcommand{\kkppz}{K^-\pi^-K^+\pi^0\nu_\tau} 
\newcommand{\tkkpi}{\tau^-\to K^-\pi^-K^+\nu_\tau} 
\newcommand{\kkpi}{K^-\pi^-K^+\nu_\tau} 
\newcommand{\tkpp}{\tau^-\to K^-\pi^+\pi^-\nu_\tau} 
 
\newcommand{\ppp}{\pi^-\pi^+\pi^-\nu_\tau}

\begin{document}

\preprint{CLNS 03-1857}       % for CLNS notes
\preprint{CLEO 03-17}         % for CLNS notes

\title{Wess-Zumino Current and the Structure of 
the Decay $\tkkpi$ }  
\author{T.~E.~Coan}
\author{Y.~S.~Gao}
\author{F.~Liu}
\author{R.~Stroynowski}
\affiliation{Southern Methodist University, Dallas, Texas 75275}
\author{M.~Artuso}
\author{C.~Boulahouache}
\author{S.~Blusk}
\author{J.~Butt}
\author{E.~Dambasuren}
\author{O.~Dorjkhaidav}
\author{J.~Haynes}
\author{N.~Menaa}
\author{R.~Mountain}
\author{H.~Muramatsu}
\author{R.~Nandakumar}
\author{R.~Redjimi}
\author{R.~Sia}
\author{T.~Skwarnicki}
\author{S.~Stone}
\author{J.C.~Wang}
\author{Kevin~Zhang}
\affiliation{Syracuse University, Syracuse, New York 13244}
% ...
%\author{(CLEO Collaboration)} %FOR PRD_SPECIAL_CHANGEME
\collaboration{CLEO Collaboration} %FOR PRL,CLNS
\noaffiliation
\begin{abstract}
We present the first study of the vector (Wess-Zumino) current in
$\tkkpi$ decay using 
data collected with the CLEO III detector at the Cornell
Electron Storage Ring. % near $\sqrt s=10.6$ GeV. 
We determine the  quantitative 
contributions to the decay width from the vector and 
axial vector currents. 
% to be $(55.7\pm8.4\pm4.9)\%$ and $(44.3\pm8.4\pm4.9)\%$,
%respectively, where the errors are statistical and systematic. 
Within the framework of a model by K$\rm\ddot u$hn and Mirkes, we identify the
quantitative contributions to the total decay rate from the intermediate 
states $\omega\pi$, $\rho^{(\prime)}\pi$ and $K^{*}K$. %to be 
%${\cal R}_{\rm WZ}^{\omega\pi}=(3.4\pm0.9\pm1.0)\%$, 
%${\cal R}_{\rm AV}^{\rho^{(\prime)}\pi}=(2.5\pm0.8\pm0.4)\%$, 
%${\cal R}_{\rm WZ}^{K^{*}K}=(60.8\pm8.5\pm6.0)\%$   
%and ${\cal R}_{\rm AV}^{K^{*}K}=(46.8\pm8.4\pm5.2)\%$, 
%respectively, where the decay to $K^*K$ can proceed via both the 
%vector (Wess-Zumino) and axial vector currents.  
\end{abstract}
\maketitle  

Hadronic $\tau$ decays provide a powerful tool to study 
low energy strong-interaction physics. The decay $\tkkpi$ 
can proceed via both the vector and axial vector currents~\cite{PRD47};  
%and, therefore, it is sensitive to
the vector current proceeds via the  Wess-Zumino mechanism~\cite{WZ}.  
This model-independent mechanism plays an important role in 
hadron dynamics. % and has been observed in tau decays~\cite{eta}. 
It is applicable to $\tau$ decays~\cite{PRD47,LI} with 
three or more hadrons in the final state and violates the 
rule that the vector and axial vector currents produce an even and an odd 
number of pseudoscalars, respectively. 
%It has been studied in $\tau^-\to\eta\pi^-\pi^0\nu_\tau$ decay~\cite{eta}. 
%allowing the vector current to produce an odd number of pseudoscalars.
%It applies to $\tau$ decays to at least three hadrons. 
Further interest in this decay is motivated by the fact 
that more precise knowledge of the $K\pi \bar K$ 
mass spectrum near the $\tau$ mass is essential to give 
more stringent constraints on the $\tau$ neutrino mass~\cite{DELCO}. 
In this Letter, we present the first study of the 
vector (Wess-Zumino) current, as well as the axial vector current, 
in the decay $\tkkpi$,
 and determine their quantitative contributions to the decay width. 
%The unbinned maximum likelihood method is used to study 
%the structure of the decay which is essential for
%exploring the constraints on the $\tau$ neutrino mass.   
Charge conjugate decays are implied throughout this Letter.

In the Standard Model, the general form for the Cabibbo-allowed semileptonic 
$\tau$ decay matrix element %$\cal{M}$ 
can be expressed as 
%\begin{eqnarray}
${\cal M}=\frac{G}{\sqrt2}V_{ud}
\bar u(q_\nu)\gamma^\mu(1-\gamma_5)u(q_\tau) J^\mu$, 
%\label{calM} 
%\end{eqnarray}
where $G$ is the Fermi coupling constant, $V_{ud}$ is a 
Cabibbo-Kobayashi-Maskawa  matrix element; 
 $q_\nu$ and $q_\tau$ are
the four-momenta of the $\tau$ neutrino and the $\tau$ lepton, 
respectively; 
$J^\mu\equiv\langle \mbox{hadrons}|V^\mu-A^\mu|0\rangle$ is the hadronic 
current, and $V^\mu$ and $A^\mu$ are the vector and axial vector
quark currents. %respectively and $q_\nu$ and $q_\tau$ are the
%four-momenta of the $\tau$ neutrino and the $\tau$ lepton, 
%The cosine and 
%the sine of the Cabibbo angle ($\theta_C$) %in Eq.~\ref{calM}
%have to be used for Cabibbo-allowed $\Delta S=0$ and Cabibbo-suppressed 
%$|\Delta S|=1$ decays, respectively. 
The most general ansatz for 
the hadronic current of three hadrons is characterized by 
four form-factors~\cite{ZPC56}   in 
terms of the hadrons' momenta $q_i$ (here, $i=1$ for $K^-$, 
$i=2$ for $\pi^-$ and $i=3$ for $K^+$), 
\begin{eqnarray} 
J^\mu&=&
\left(q_1^\mu-q_3^\mu-Q^\mu\frac{Q(q_1-q_3)}{Q^2}\right)F_1(s_1,s_2,Q^2)
\nonumber\\
&&+ \left(q_2^\mu-q_3^\mu-Q^\mu\frac{Q(q_2-q_3)}{Q^2}\right)F_2(s_1,s_2,Q^2)
\nonumber\\
&&+
i\epsilon^{\mu\alpha\beta\gamma}q_{1\alpha}q_{2\beta}q_{3\gamma} 
F_3(s_1,s_2,Q^2) \nonumber\\&&  
 +Q^\mu F_4(s_1,s_2,Q^2),
\end{eqnarray}
where $Q^\mu=q_1^\mu+q_2^\mu+q_3^\mu$. 
All four form-factors  $F_1$ through $F_4$ are %model-dependent 
functions of $Q^2=(q_1+q_2+q_3)^2$, $s_1=(q_2+q_3)^2$, 
$s_2=(q_1+q_3)^2$ and $s_3=(q_1+q_2)^2$. 
The terms proportional to $F_1$ 
and $F_2$ originate from the axial vector current ($J^P=1^+$).  
The  vector current ($J^P=1^-$) originating from the Wess-Zumino mechanism 
gives rise to the term proportional  to $F_3$. 
$F_4$ is due to the spin-zero scalar current 
($J^P=0^+$) that is 
expected to be small~\cite{PRD47,ZPC69}  
and in this analysis it will be set to zero, 
although we will consider its contribution to the systematic uncertainties  
later.  

Tau decay to three mesons is conveniently analyzed in the 
hadronic rest frame, where ${\bf q_1}+$ $\bf{q_2}+{\bf q_3}={\bf 0}$. 
%The orientation of the hadronic
%system is characterized by three Euler angles~\cite{ZPC56}. One of them, 
The angle $\beta$~\cite{ZPC56} is defined as the angle between the direction of 
the hadronic system
in the laboratory frame  and the normal to the plane defined by the momenta 
of particles 1 and 2,
\begin{eqnarray}
\cos\beta=-\hat {\bf Q}\cdot\mbox{norm}(\hat{\bf q}_1\times\hat{\bf q}_2).  
\end{eqnarray} 
The angle $\theta$~\cite{ZPC56}, defined to be the angle between the flight direction 
of the $\tau$ lepton in the laboratory frame 
%(=direction %of the $\tau$ polarization) 
and the direction of the hadronic system  as seen 
in the $\tau$ rest frame, is related to the energy $E_h$ of the 
hadronic system in the laboratory  frame by   
\begin{eqnarray}
\cos\theta=\frac{2xm_\tau^2-m_\tau^2-Q^2}
{(m_\tau^2-Q^2)\sqrt{1-4m_\tau^2/s}},
\label{costheta} 
\end{eqnarray} 
where $x={2E_h}/{\sqrt s}$ and $s$ is the square of the center of mass energy. 
The angle $\psi$~\cite{ZPC56} between the flight direction of the $\tau$ lepton 
and that of the laboratory frame  
as seen from the hadronic rest frame, $-\hat{\bf Q}$, is given by 
\begin{eqnarray}
\cos\psi=\frac{x(m_\tau^2+Q^2)-2Q^2}{(m_\tau^2-Q^2)\sqrt{x^2-4Q^2/s}}.
\label{cospsi}
\end{eqnarray} 
%The angle $\beta$ together with other two Euler angles~\cite{ZPC56}, 
%characterizes the orientation of the 
%hadronic system, and the angles $\theta$  and  $\psi$ the relative orientation
% of the laboratory frame, hadronic system and $\tau$ rest frame.   

The differential decay width 
for the decay $\tkkpi$ with the scalar 
%for the Cabibbo-allowed tau decay to three hadrons with the scalar 
contribution neglected, after integrating over the 
unobserved neutrino direction, is given~\cite{ZPC56} by  
%\begin{widetext} 
\begin{eqnarray}
{d\Gamma(\tau\to K\pi\bar K\nu_\tau)
\over dQ^2ds_1ds_2}&=&
\frac{G^2}{12m_\tau}%\left(g_V^2+g_A^2\right)
\left|V_{ud}\right|^2
\frac{1}{(4\pi)^5}\frac{(m_\tau^2-Q^2)^2}{Q^4}
\nonumber \\ && %\times
{ \left\{\left(1+\frac{2Q^2}{m_\tau^2}\right)
\left(W_A+W_B\right)%+3W_{SA}
\right\}.}
\label{XQ2}
\end{eqnarray}
%\end{widetext} 
Here, the structure functions $W_A$ and $W_B$ can be %in general 
expressed in terms of the form factors as: 
\begin{eqnarray}
W_A&=&(x_1^2+x_3^2)|F_1|^2+(x_2^2+x_3^2)|F_2|^2\nonumber\\ &&
+2(x_1x_2-x_3^2){\rm Re}(F_1F_2^*),\nonumber\\ 
W_B&=&x_4^2|F_3|^2,
\label{Ws}
\end{eqnarray}
where $x_i$~\cite{ZPC56} are known functions of $Q^2$, $s_1$, $s_2$ and $s_3$. 
{Thus $W_A$ and $W_B$ govern the rate and 
the distributions of $Q^2$, $s_1$, $s_2$ and $s_3$.
% $s_1$, $s_2$, $s_3$ and $Q^2$,  
There is no interference between the axial vector current 
contributions ($F_1, F_2$) and the vector current contribution
($F_3$) to the decay width}. 
% which in turn helps simplify the fitting procedure
% based on limited statistics. %us 
% to determine the anomaly contribution to the total decay width without
%extracting 16 structure functions. 

A parameterization of the form-factors  
%for the $\tau^-\to K^-K^+\pi^-\nu_\tau$ decay 
is given in~\cite{ZPC69}, based on chiral perturbation theory at low momenta
and meson resonance dominance at higher momenta. 
It has been implemented in the Monte Carlo program
{\sc koralb}~\cite{korb} as: 
%\begin{widetext}
\begin{eqnarray}
F_{1}%(Q^2,s_2,s_3)
                   &=&%-\frac{\sqrt 2}{3f_\pi}
                   %\mbox{BW}_{a_1}(Q^2) \trhozm(s_2)=
                   -\frac{\sqrt 2}{3f_\pi}\mbox{BW}_{a_1}(Q^2)
                   \frac{\mbox{BW}_\rho(s_2) + \beta_\rho
                    \mbox{BW}_{\rho'}(s_2) }{1 + \beta_\rho}, \nonumber\\
F_{2}%(Q^2,s_1,s_3)
                   &=&%-\frac{\sqrt 2}{3f_\pi}
                    %\mbox{BW}_{a_1}(Q^2) \tkszm (s_1)=
                  -\frac{\sqrt2}{3f_\pi}\cdot R_F\cdot
                   \mbox{BW}_{a_1}(Q^2)\cdot\mbox{BW}_{K^*}(s_1), \nonumber \\ 
F_{3}%(Q^2,s_1,s_2)
                 &=&-{1\over 2\sqrt 2\pi^2f^3_\pi}\cdot\sqrt{R_B}
                 \cdot\frac{\mbox{BW}_{\omega}(s_2)+\alpha 
                 \mbox{BW}_{K^*}(s_1) }{1+\alpha} 
                 \nonumber \\ &&\cdot    
                 \frac{\mbox{BW}_\rho(Q^2) + \lambda \mbox{BW}_{\rho'}(Q^2)
                       +\delta \mbox{BW}_{\rho''}(Q^2)}
                  {1+\lambda+\delta},
\label{simple} 
\end{eqnarray}
%\end{widetext}
where $f_\pi=0.0933$ GeV and $\beta_\rho=-0.145$~\cite{ZPC69,korb}. 
$\mbox{BW}_{X}(s)$ is the two- (dependent on $s_1$ or $s_2$) or three-
(dependent on $Q^2$) particle Breit-Wigner propagator with an energy 
dependent width. 
The CLEO result on the $a_1$ parameterization derived 
from $\tau^-\to\pi^-\pi^0\pi^0\nu_\tau$ decay~\cite{pi2pi0} is  used. 
The parameterizations of two-particle 
Breit-Wigner functions take the forms used in~\cite{ZPC69}.  
The parameters used for the three-particle Breit-Wigner functions
are taken from the Particle Data Tables~\cite{PDG}: $m_\rho=0.770$ GeV,
 $\Gamma_\rho=0.151$ GeV,   $m_{\rho'}=1.465$ GeV,  
$\Gamma_{\rho'}=0.310$ GeV, $m_{\rho''}=1.700$ GeV and  
$\Gamma_{\rho''}=0.240$ GeV. 
%\begin{eqnarray}
%m_\rho = 0.770 \mbox{ GeV},    &&\Gamma_\rho = 0.151 \mbox{ GeV},\nonumber \\
%m_{\rho'} = 1.465 \mbox{ GeV}, &&\Gamma_{\rho'} = 0.310 \mbox{ GeV},\nonumber\\
%m_{\rho''} = 1.700 \mbox{ GeV}, &&\Gamma_{\rho''} = 0.240 \mbox{ GeV}.
%\label{masses}
%\end{eqnarray}

The form-factors  $F_1$ and $F_2$ correspond to the %contribution from 
axial vector current processes $a_1\to\rho^{(\prime)}\pi^-$ with 
$\rho^{(\prime)}\to K^-K^+$ and $a_1\to K^*K^-$ with $K^*\to K^+\pi^-$,
respectively. The form-factor $F_3$ represents the %contributions from the 
vector current processes 
$\rho^{(\prime,\prime\prime)}$ decays to $K^*K^-$
and $\omega\pi^-$ with $K^*\to K^+\pi^-$ and $\omega\to K^-K^+$.   
$G$ parity conservation forbids the %, the vector resonance $V$ in the 
vector current process 
$W^-\to\rho^{(\prime,\prime\prime)}\to\rho^{(\prime)}\pi^-$ with 
$\rho^{(\prime)}\to K^-K^+$. Instead, in~\cite{ZPC69} the process 
$W^-\to\rho^{(\prime,\prime\prime)}\to\omega\pi^-$  is considered, although the 
subsequent decay $\omega\to K^-K^+$ proceeds only through the 
high-mass tail of its Breit Wigner propagator.
We make use of this model in this analysis, 
although we will be unable to distinguish whether 
the decay proceeds via the $\omega$ meson or via some other
resonance with pole mass outside the kinematic limits
for the $KK$ mass.
%Because of the antisymmetric 
%tensor structure of the Wess-Zumino current, the $K^-K^+$ must 
%system must have positive $G$ parity. %, thus the vector 
%%$V=\rho^{(\prime)}$ is not allowed in the Wess-Zumino current, 
%The two kaon resonance can not be a $\rho^{(\prime)}$ resonance  
%in the vector (Wess-Zumino) current. 
%Only the lightest vector meson $\omega$ qualifies~\cite{ZPC69}. The 
%data are insensitive to the nature of the $\omega$ resonance, only
%to the presence of the tail of its Breit-Wigner function. 

%We observe clear discrepancies in the mass distributions~\cite{tau02} 
% between the  data and the default version of {\sc koralb}~\cite{korb} 
%which show that the decay is not well-modelled. 
%The parameter $\beta_\rho$ has been well-determined using both 
%$e^+e^-$ and $\tau$ data~\cite{ZPC69}. 
%Besides the other  model parameters $\alpha$, $\lambda$ and $\nu$ 
%in Eq.~\ref{simple},   
%we introduce two more parameters $R_B$ and $R_F$, defining the relative 
%strenghts of $W_B$ and $W_A$ and of $F_2$ and $F_1$, that is, $W_B$ in
% Eq.~\ref{XQ2} and $F_2$ in Eq.~\ref{Ws} are replaced with 
% $W_B\to R_B W_B$ and $F_2\to R_F F_2$ (in fact $R_F=1$ and $R_B=1$ 
%in the default version of {\sc koralb}). These five parameters  
%will be determined in this analysis by simultaneously fitting
%the invariant mass spectra of $K^-K^+\pi^-$, $K^+\pi^-$ and $K^-K^+$. 

The default version of {\sc koralb}~\cite{korb} does not model 
the decay $\tkkpi$ well, as indicated by the discrepancies in describing
the invariant mass distributions~\cite{tau02}. To improve the agreement,
we introduce two more parameters $R_B$ and $R_F$, describing the 
relative strengths of $W_B$ and $W_A$ and of $F_2$ and $F_1$ in Eqs.~\ref{XQ2}
and~\ref{Ws}. These two parameters are explicitly expressed in
our generalization of the form-factors in Eq.~\ref{simple}. The five model 
parameters: $\alpha$, $\lambda$, $\delta$, $R_B$ and $R_F$ are determined 
in this analysis by a simultaneous fit to the invariant mass  spectra of 
$K^-\pi^-K^+$, $K^+\pi^-$ and $K^-K^+$. These five parameters 
allow for a fit that properly describes all the mass spectra.  In general, 
these parameters should be complex, allowing additional interferences. 
However, due to limited statistics, we  have set them real and taken into
account only the inherent phase of the Breit-Wigner functions in 
Eq.~\ref{simple}.

%\section{Fit Results} 
The data sample used for this analysis corresponds to
an integrated luminosity of 7.77~fb$^{-1}$ taken
on or near the $\Upsilon(4S)$.  It contains  about 7.09$\times10^6$
tau-pairs. The data  
 were collected with the CLEO~III 
detector~\cite{CLEOIII} located at the $e^+e^-$ 
Cornell Electron Storage Ring.  
%The data sample consists of 7.77~fb$^{-1}$ taken 
%on or near the $\Upsilon(4S)$,  corresponding to 7.09$\times10^6$ 
%$\tau^+\tau^-$ pairs. 
The CLEO~III detector configuration %bkh
features a four-layer silicon strip 
vertex detector, a wire drift chamber, and,
most importantly for this analysis, a ring 
imaging Cherenkov (RICH) particle identification system.
A detailed description of the RICH performance 
can be found in~\cite{RICH}. 
%The response of the detector 
%is studied with a detailed {\sc geant}-based~\cite{GEANT} simulation.  

To select the three-prong $\tau$ decays to the $\kkpi$ final state, 
we use one-prong decays of the other $\tau$ to $e^+\nu\nu$, 
$\mu^+\nu\nu$, $\rho^+\nu$ and  $\pi^+(K^+)\nu$ to tag tau-pair events. 
% from the opposite-side as a tag. 
Pion and kaon identification is obtained by a combination of 
the RICH information 
with $dE/dx$ measured in the drift chamber. 
% to provide good pion and kaon separation for the three-prong side. 
We use the event selection criteria listed in~\cite{PRL}  
 and find 2255 candidate signal events in our data sample. 
%Detailed descriptions of the event selection and of 
%the pion and kaon separation can be found in~\cite{PRL}. 
%In that paper, we concentrated on the measurements
%of the branching fractions for the $\tau$ decays to three charged hadrons.  
The branching fraction obtained using this data sample for the decay $\tkkpi$ 
is consistent with the previously published result~\cite{PRL}.  
%which presents the measurements of the branching fractions for 
%the tau decays to three charged hadrons.     

Background contributions and efficiencies are obtained 
using Monte Carlo events from the {\sc koralb}~\cite{korb} 
and {\sc jetset}~\cite{jetset} generators for %($q\bar q\to$ hadrons) 
the tau-pair production and $q\bar q\to$ hadron processes, respectively. 
These events are then 
processed by the {\sc geant}-based~\cite{GEANT} CLEO detector simulation
and pattern reconstruction.
%passed through the {\sc geant}-based~\cite{GEANT} CLEO  detector simulation. 
In our sample, $256\pm16\pm46$ events are attributed  
to the backgrounds from the continuum $q\bar q$  production 
and from tau decays to other channels. 
%The errors are statistical and systematic, respectively. 
The dominant backgrounds are  due to 
pion mis-identification or  missing $\pi^0$ from 
the tau cross-feed decays $\tkpp$, $\ppp$, and $\kkppz$, % due to pion,  
%mis-identification or  missing $\pi^0$, 
or from $\tkkpi$ decay
 with a  mis-identified tag side. % side was mis-identified.   
Each of these sources contributes about 20\% of the total backgrounds. 
The events with the three-prong side identified as $\kkpi$ and  
the tag side mis-identified are analysed in the same way as the signals.
% which results in 2.3\% higher efficiency.  

%The reconstruction efficiencies show a slight dependence on the    
%$K^-K^+\pi^-$, $K^+\pi^-$ and $K^-K^+$ masses. 
%The maximum variation  
%in efficiencies across the mass distributions is predicted from
%the simulation to be of order 10\%. 

We use the unbinned extended maximum likelihood method to extract the 
parameters of the hadronic structure of the decay described above. 
% to study the hadronic structure of the decay. 
%by fitting  the invariant mass spectra
%of $K^-K^+\pi^-$, $K^+\pi^-$ and $K^-K^+$ simultaneously. 
The likelihood function, $\cal L$, has the form
%\begin{widetext} 
%\begin{eqnarray}
%{\cal L}=\frac{e^{-m}m^n}{n!}\prod_{i=1}^{n}
%\frac{N^S\cdot{\cal P}_i^S(\alpha,\lambda,\delta,R_B,R_F,
%Q^2,s_1,s_2)+N^B\cdot{\cal P}^B_i(\alpha,\lambda,\delta,R_B,R_F,Q^2,s_1,s_2)}{N^S+N^B}.  
%\end{eqnarray}
%\end{widetext} 
\begin{eqnarray}
{\cal L}&=&\frac{e^{-m}m^n}{n!}\prod_{i=1}^{n}
\frac{1}{n}
\left\{N^S{\cal P}_i^S(\alpha,\lambda,\delta,R_B,R_F,
Q^2,s_1,s_2)\right.\nonumber\\
&&\left.~~~~~~~~+N^B{\cal P}^B_i(\alpha,\lambda,\delta,R_B,R_F,Q^2,s_1,s_2)\right\}. 
\end{eqnarray}
The number of signal and background events are denoted by 
$N^S$ and $N^B$ with $n=N^S+N^B$. $m$ is the number of the expected 
events.  ${\cal P}_i^S$ and ${\cal P}_i^B$ represent   
the normalized probability density functions (PDF) for event $i$ to be 
either a signal or a background event. 
The signal PDF is  formed from the product of 
one-dimensional PDFs %(functions of $s_1$ or $s_2$ or $Q^2$ only) 
obtained from integrations over
the differential decay width described by Eq.~\ref{XQ2}, convoluted with the 
mass-dependent efficiencies. 
%The maximum variation  
%in efficiencies across the mass distributions is predicted from
%the simulation to be of order 10\%. 
The background PDFs are obtained from
Monte Carlo simulations. %samples discussed above. 
The fitting procedure has been 
tested with Monte Carlo samples  to verify its performance. 

\begin{figure}[htbp]
\center{\mbox{\includegraphics[width=1.0\textwidth]{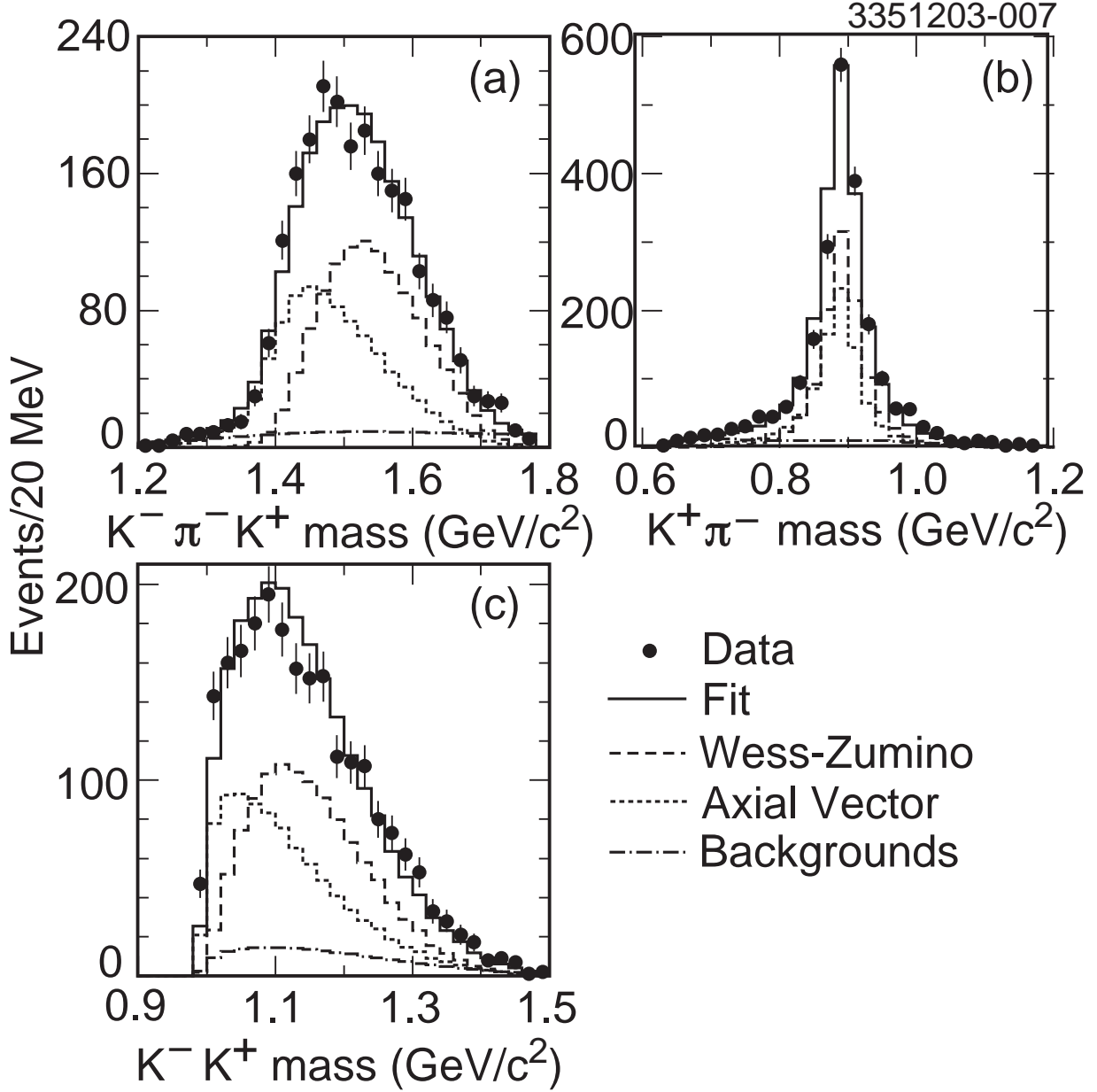}}}
\caption{Projections of the fits (full lines) 
onto the a) $K^-\pi^-K^+$, b) $K^+\pi^-$ and c) 
$K^-K^+$ mass axes from the data (dots with error bars) superimposed 
with the contributions from the vector (Wess-Zumino) current (dashed) and 
the axial vector current (dotted) and all backgrounds (dot-dashed).
%d) the distributions of the $K^+$  helicity angle in the 
%$K^{*0}\to K^+\pi^-$ rest frame evaluated in the three hadrons rest frame
%with the hadronic mass less (long-dotted-dashed) and greater
% (long-dashed lines) than 1.5 GeV$/c^2$, see the text.  
%The dots represent all data points, the dashed lines all backgrounds.
}
\label{projdata}
\end{figure}
 
\begin{figure}[htbp]
\center{\mbox{\includegraphics[width=1.0\textwidth]{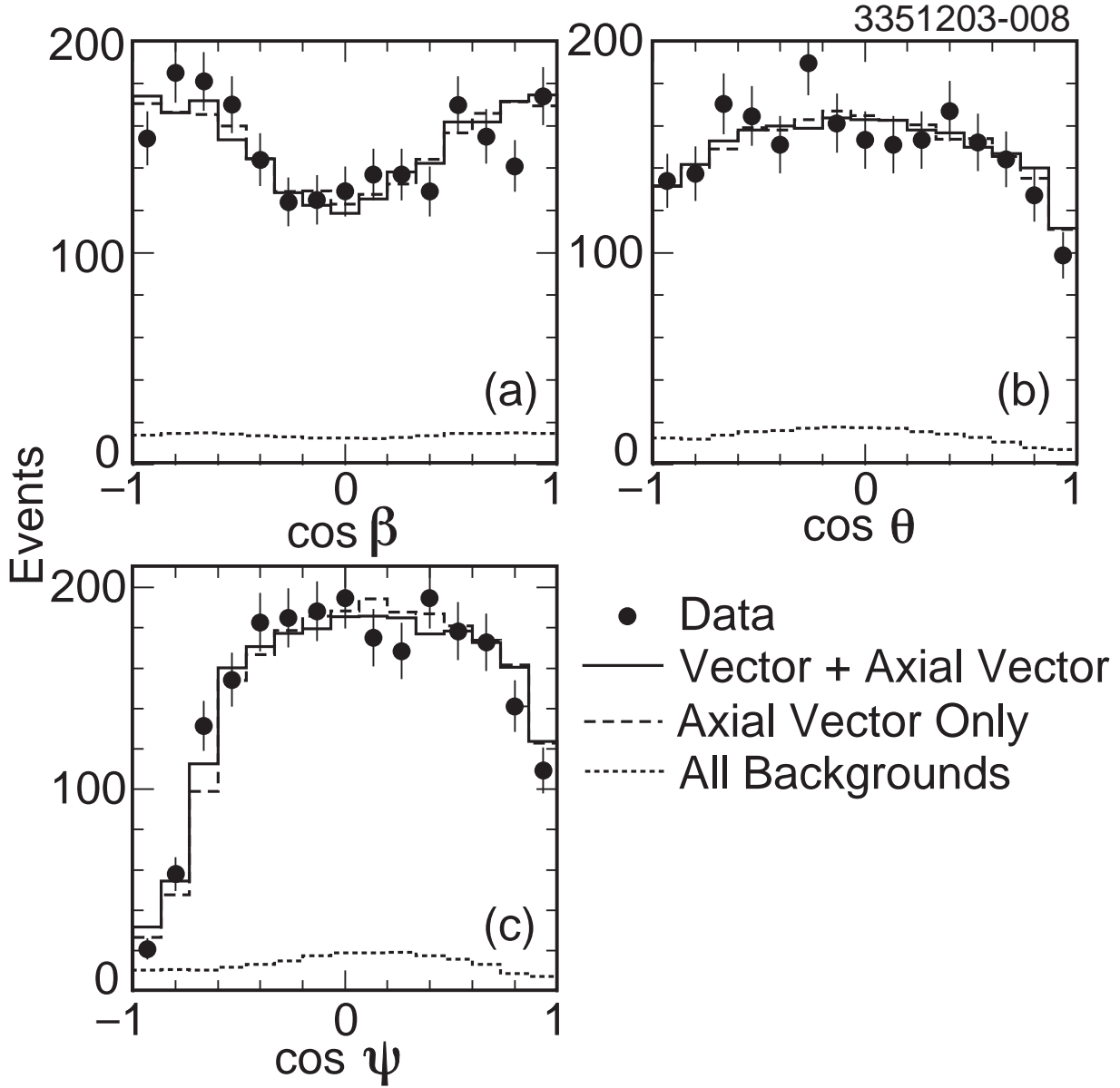}}}
\caption{Angular distributions of a) $\cos\beta$, b) $\cos\theta$ 
and c) $\cos\psi$ for the data with statistical errors only. 
%Solid lines present the normalized Monte Carlo sample 
% generated with both the vector and axial vector current 
%contributions using the parameters from the fits. Dashed lines 
%describe the normalized Monte Carlo sample 
%generated with the axial vector contribution only.  
Solid lines represent our model of the vector (Wess-Zumino) and axial-vector
signal contributions and all backgrounds, using the parameters
from the full fit to the data.
Dashed lines represent our model of signal and backgrounds
with the vector (Wess-Zumino) contribution excluded.
Dotted lines show all backgrounds.   } 
\label{cosall}
\end{figure}
 
The results of the unbinned extended maximum likelihood fits are projected onto
the $K^-\pi^-K^+$, $K^+\pi^-$ and $K^-K^+$ mass axes with different 
contributions superimposed as illustrated in Fig.~\ref{projdata}.  
%The fits to the $K^-K^+\pi^-$, $K^+\pi^-$ and $K^-K^+$ mass spectra  
%from the data  are shown in Fig.~\ref{projdata} with different components
%overlaid. 
We observe that about half of the $\tau$ decay to $\kkpi$ proceeds via
the vector (Wess-Zumino) current.
The fit parameters are $\alpha=0.471\pm0.060$, $\lambda=-0.314\pm0.073$,
$\delta=0.101\pm0.020$, $R_{B}=3.23\pm0.26$ and 
$R_F=0.98\pm0.15$ (statistical errors only). 
%The fits show that 
%there is significant contribution to the decay rate 
%from the Wess-Zumino vector current. 
The angular distributions of 
$\cos\beta$, $\cos\theta$ and $\cos\psi$ for the data  and for the two Monte 
Carlo samples generated using the parameters from the fits 
%with and without the contribution from the vector (Wess-Zumino) current 
are shown in Fig.~\ref{cosall}. %As can be seen, 
The  comparison shows that
the Monte Carlo samples with and without
the contribution from the vector (Wess-Zumino) current both  describe the 
data well within the current statistics; %That means  that 
the angular observables  therefore have little sensitivity to the 
presence of the vector (Wess-Zumino) current in our sample.  Our ability to 
distinguish the axial vector contribution from the vector (Wess-Zumino) 
contribution
comes from the invariant mass distributions, and thus 
depends on the correct modeling of the associated form-factors, as 
constrained by our data. 

%There are some sources of systematic error which may affect the fit results. 
We have considered the following sources of systematic errors: 
the systematic errors for the five model parameters and for 
the fractions of the vector and axial vector current contributions to 
the decay rate (including those of the intermediate states) 
come from the mass dependence of the detection efficiency,
the uncertainties in the modeling of the mass resolutions, 
%the uncertainties in the background level and PDFs, 
the uncertainties in the parameterizations
of the resonances and from the neglected contribution of the 
scalar current. 
To estimate the effects of the detection efficiencies, we introduce an
artificial linear mass dependence of the efficiencies and 
vary the corresponding slopes by $\pm10\%$.
%the slopes of the linear efficiency functions are varied by $10\%$.  
%Changes in the results are taken as the systematic errors and 
%found to be 10\% (relative) for parameter $R_B$ and less than 4\% for 
%other parameters and six ratios to be presented. 
We estimate the associated systematic error on the relative strength
of the vector and axial vector currents as $\Delta R_B/R_B=10\%$.
The corresponding systematic errors on all other fit parameters 
and the fractions of the vector and axial vector current contributions 
to the decay rate %be presented 
are less than 4\% of their values. 
The typical mass resolutions of the hadronic final state in  the decay $\tkkpi$
 are
about 2--3 MeV, i.e., much smaller than the 
widths of the resonances considered. %Compared to the wide resonances 
%($\Gamma>50$ MeV/c$^2$) which dominate the structure, 
The uncertainty in the modeling of the detector resolutions 
%makes only 
contributes a $<$2\% systematic error. 
%the detector smearing effects therefore contribute only 
%a small part of the systematics.
The  uncertainty introduced by the neglected  scalar current 
%from the scalar part of the axial vector which is set to zero 
is estimated at the model predicted level~\cite{ZPC69,korb} and 
found to be less than 1\%.   
%due  to its very small contribution to the decay width~\cite{ZPC69}.  
In addition, when we vary the normalization  and the shapes of 
the background PDFs by $\pm1\sigma$,  the fit results vary between 1\% and 
12\%. When the parameter $\beta_\rho$ in Eq.~\ref{simple} varied 
 by $\pm100\%$,  the results change 37\% for the parameter $R_F$ and 
less than 3\% for all other fit parameters and the contributions 
of the vector and axial vector currents.   
The dominant uncertainty in the fit results 
arises from the errors of the parameterizations of the poorly measured 
$\rho'$ and $\rho''$ resonances~\cite{PDG}.
% when their masses and widths are varied by $\pm1\sigma$.
 The corresponding   
contribution to the overall systematic uncertainties for 
the parameters  $\alpha$, $\lambda$, $\delta$, $R_B$,  and $R_F$
is 5\%, 23\%, 153\%, 56\%, 1\%, 
%for the fit parameters $\alpha$, $\lambda$, $\delta$, $R_B$,  and $R_F$, 
respectively. Similarly it is 8\%, 10\%, 11\%, 12\%, 9\% and 27\% 
for the fractions ${\cal R}_{\rm WZ}$,  
${\cal R}_{\rm AV}$, ${\cal R}_{\rm AV}^{K^*K}$, 
 ${\cal R}_{\rm AV}^{\rho^{(\prime)}\pi}$, 
${\cal R}_{\rm WZ}^{K^{*}K}$
and ${\cal R}_{\rm WZ}^{\omega\pi}$ to be presented. 
%All five parameters are correlated.  
%The contribution from each of these sources varies among the 
%measured quantities. % their maximum contributions
%(relative) to the overall systematic errors vary from about 4\% to 110\%'.  
 %which are not well established from the PDG~\cite{PDG}. %The total 
%sysmtematic error is quadratic sum over individuals with the larger one 
%used if two are available in cases of $+\sigma$ and $-\sigma$ variations. 
%All of the above sources of systematic error are combined in quadrature. 
We add all estimates of the systematic errors in quadrature and obtain 
the overall systematic uncertainties for the five parameters 
$\alpha$ (7\%), $\lambda$ (25\%), $\delta$~(154\%), $R_B$ (59\%), 
and $R_F$ (37\%) and for the fractions
${\cal R}_{\rm WZ}$ (9\%), ${\cal R}_{\rm AV}$ (11\%), 
${\cal R}_{\rm AV}^{K^*K}$ (11\%),
 ${\cal R}_{\rm AV}^{\rho^{(\prime)}\pi}$ (16\%),
${\cal R}_{\rm WZ}^{K^{*}K}$ (10\%)
and ${\cal R}_{\rm WZ}^{\omega\pi}$ (29\%). 
There is a large uncertainty in the contribution of the
$\rho''$ resonance to the vector current.
This has little effect on the total vector contribution
because it is strongly suppressed by phase space,
and its contribution is easily absorbed into the 
contribution from the $\rho'$ resonance.

%\section{Summary}
In summary, 
we have presented the first study of the vector (Wess-Zumino) current in
the decay $\tkkpi$ and determined its contribution, as well as that from 
the axial vector current, to be:  
\begin{eqnarray}
{\cal R}_{\rm WZ}&=&\frac{\Gamma_{\rm WZ}}{\Gamma_{\rm tot}}=(55.7\pm8.4\pm4.9)\%,\nonumber\\ 
{\cal R}_{\rm AV}&=&\frac{\Gamma_{\rm AV}}{\Gamma_{\rm tot}}=(44.3\pm8.4\pm4.9)\%.
\end{eqnarray} 
The errors are statistical and systematic, respectively. 

The structure of the decay $\tkkpi$  can be modeled using  
the parameterization given by Eqs.~\ref{XQ2}--\ref{simple}, 
%in~\cite{ZPC69} and~\cite{korb}, 
with the parameters determined to be: 
$\alpha=0.471\pm0.060\pm0.034$, $\lambda=-0.314\pm0.073\pm0.080$,
$\delta=0.101\pm0.020\pm0.156$, $R_B=3.23\pm0.26\pm1.90$, 
and $R_F=0.98\pm0.15\pm0.36$. 
%\begin{eqnarray}  
%\alpha&=&0.471\pm0.060\pm0.034,\nonumber\\ 
%\lambda&=&-0.314\pm0.073\pm0.080,\nonumber \\
%\delta&=&0.101\pm0.020\pm0.156,\nonumber \\ 
%R_B&=&3.23\pm0.26\pm1.90,\nonumber \\ 
%R_F&=&0.98\pm0.15\pm0.36. 
%\end{eqnarray} 

In the context of the model by K$\rm\ddot u$hn and Mirkes~\cite{ZPC56}, 
where the axial vector current proceeds via 
$a_1\to\rho^{(\prime)}\pi^-$ and $a_1\to K^*K^-$, 
and the vector (Wess-Zumino) current via 
$\rho^{(\prime,\prime\prime)}\to K^*K^-$ and $\omega\pi^-$,   
we can also separate the  individual contributions from the intermediate 
states $\omega\pi^-$,  $\rho^{(\prime)}\pi^-$ and $K^{*}K^-$  
to the total decay width as 
\begin{eqnarray}
{\cal R}_{\rm WZ}^{\omega\pi}&=&\frac{\Gamma_{\rm WZ}^{\omega\pi}}{\Gamma_{\rm tot}}
                 ~~=(3.4\pm0.9\pm1.0)\%, \nonumber \\
{\cal R}_{\rm AV}^{\rho^{(\prime)}\pi}&=&\frac{\Gamma_{\rm AV}^{\rho^{(\prime)}\pi}}{\Gamma_{\rm tot}}
                 ~ =(2.5\pm0.8\pm0.4)\%, \nonumber \\ 
{\cal R}_{\rm WZ}^{K^{*}K}&=&\frac{\Gamma_{\rm WZ}^{K^*K}}{\Gamma_{\rm tot}}
                  =(60.8\pm8.5\pm6.0)\%, \nonumber \\  
{\cal R}_{\rm AV}^{K^{*}K}&=&\frac{\Gamma_{\rm AV}^{K^*K}}{\Gamma_{\rm tot}}
                  =(46.8\pm8.4\pm5.2)\%. 
\end{eqnarray}
These fractions do  not add up to 100\% due to  the 
interference of the amplitudes of  the intermediate states. 
The decay is dominated by the vector (Wess-Zumino) and axial vector 
current processes via the intermediate state $K^*K$. 
The ratio $R_{\rm AV}^{K^*K}$, together with ${\cal B}(\tkkpi)$~\cite{PRL},
yields ${\cal B}({a_1\to K^*K})=(2.2\pm0.5)\%$ which is 
compatible with the result 
${\cal B}({a_1\to K^*K})=(3.3\pm0.5)\%$
extracted from the decay $\tau^-\to\pi^-\pi^0\pi^0\nu_\tau$~\cite{pi2pi0}  
with only one threshold ($K^*K$) considered. 
The apparent shortfall of the $\pi^-\pi^0\pi^0$ mass spectrum in the previous
analysis is attributed to the neglected 
contribution from the vector (Wess-Zumino) current. 
%The CLEO result ${\cal B}(\tau^-\to a_1\nu_\tau\to [K^*K]^-\nu_\tau)= 
%(0.16\pm0.03)\%$, compared to ${\cal B}(\tau^-\to [K^*K]^-\nu_\tau)=
%(0.42\pm0.08)\%$ in the PDG~\cite{PDG}, also favors  
%more vector (Wess-Zumino) current component in this mode. 
The axial vector current contribution determined in this 
Letter is much lower than the result from ALEPH~\cite{ALEPH} 
based on CVC and limited $e^+e^-$ data. 
As noted above, we can not distinguish the 
$\rho^{(\prime,\prime\prime)}\to\omega\pi^-$ with $\omega\to K^-K^+$ 
decay from other models, such as other vector meson resonances or even 
a simple constant.  
Setting the $\omega\pi^-$ contribution to zero, we obtain 
${\cal R}_{\rm WZ}^{(K^*K)}=(50.8\pm7.7)\%$, 
${\cal R}_{\rm AV}=(49.2\pm7.7)\%$, 
${\cal R}_{\rm AV}^{\rho^{(\prime)}\pi}=(4.5\pm1.4)\%$, and 
${\cal R}_{\rm AV}^{K^{*}K}=(51.6\pm7.8)\%$. 
However, the fit favors the presence of an additional resonance component, 
which can be associated, within the context of the model in~\cite{ZPC69}, 
with $\omega\pi^-$, over its absence by nearly 6$\sigma$  
(statistical error only). 
The model parameters presented here can reduce the model dependent 
uncertainties on the tau neutrino ($\nu_\tau$) mass measurements 
using the decay $\tkkpi$. 

We gratefully acknowledge the effort of the CESR staff 
in providing us with
excellent luminosity and running conditions.
This work was supported by 
the National Science Foundation,
the U.S. Department of Energy,
the Research Corporation,
and the 
Texas Advanced Research Program.

\end{document}